\documentstyle[twocolumn,aps,psfig,floats]{revtex}

\begin{document}
\draft                   

\title{
     \vskip -0.5in\hfill\hfil{\rm\normalsize }
     \vskip 0.4in
       Paradoxical Magnetic Cooling in a Structural Transition Model}

\author{
   Peter Borrmann\cite{Borrmannhome},   
   Heinrich Stamerjohanns, Eberhard R. Hilf}

\address{
   Department of Physics, Carl von Ossietzky University, \\
   D-26111 Oldenburg, Germany }

\author{David Tom\'anek }

\address{
Department of Physics and Astronomy, Michigan State University, \\
East Lansing, Michigan 48824-1116, USA}

\date{Received \hspace{3.0cm} }
\maketitle

\begin{abstract}
In contrast to the experimentally widely used isentropic
demagnetization process for cooling to ultra-low temperatures we
examine a particular classical model system that does not cool, but
rather heats up with isentropic demagnetization. This system consists
of several magnetite particles in a colloidal suspension, and shows
the uncommon behavior of disordering structurally while ordering
magnetically in an increasing magnetic field. For a six-particle
system, we report an uncommon structural transition from a ring to a
chain as a function of magnetic field and temperature.
\end{abstract}


Cooling to ultra-low temperatures is presently experimentally
achieved using the isentropic demagnetization process suggested in
1926 by Debye and Giauque \cite{Debye,Giauque}. Here we study the
thermal and magnetic properties of a particular model system
consisting of six magnetite nanoparticles in a colloidal suspension.
Our calculations indicate that changing the temperature or magnetic
field leads to a transition from a ring-like to a chain-like
structure. We demonstrate that such a model system, that could be
realized in a ferrofluid, would not cool but rather heat up during an
isentropic demagnetization. The system of magnetic nanoparticles
shows the uncommon behavior of disordering structurally, i.e.
increasing its volume in phase space, while ordering magnetically in
an increasing magnetic field. This behavior, associated with the
break-up of the relatively rigid rings to floppy chains, occurs once
the energy gain upon aligning a chain of dipoles with the field
exceeds the energy cost of breaking up a ring. Thus systems that
disorder structurally in high fields may be used as coolants based on
isentropic magnetization instead of demagnetization. So far, only
ferrofluid systems with a large number of particles have been
discussed as candidates for use in magneto-caloric heat engines
\cite{Rosensweig}. Here we report a paradoxical magnetic cooling
phenomenon that is unique to systems with only few particles.

Both the conventional and paradoxical process utilize the energy
change associated with particular structural changes for cooling.
The {\sl conventional} isentropic demagnetization process uses the
fact that a magnetic system, such as a spin system, orders
magnetically and thus lowers its entropy in presence of an external
magnetic field. Removal of the external magnetic field at constant
temperature leads to an increase in entropy due to magnetic
disordering, which requires energy. Decreasing the external magnetic
field at constant entropy consequently leads to a temperature
lowering. With this method, systems such as copper have been cooled
down to temperatures as low as  $50$~nK \cite{Ehnholm}.

In the following we address model systems consisting of a finite
number of magnetic particles that may undergo structural transitions.
Both the structural and magnetic degrees of freedom are important in
this system and can not be decoupled. Applying a sufficiently high
magnetic field causes the system to order magnetically while
disordering structurally, at the cost of internal energy.
Consequently, such a system will exhibit the paradoxical phenomenon of
cooling by isentropic magnetization.

The transformation from a ring to a chain is associated with freeing
up a structural degree of freedom, with a corresponding increase in
entropy. Such a transformation can be induced by a magnetic field in
a system of magnetic dipoles, where the energetics is governed by
dipole-dipole interactions between the particles and an interaction
of each particle with the external field. In small fields, the ring
is stabilized with respect to the chain structure if the gain in
dipole-dipole interaction upon connecting the chain ends
energetically outweighs the dipole misalignment energy in a bent
structure. The energy gain upon aligning all individual dipoles with
a sufficiently high applied field, on the other hand, stabilizes the
chain structure.

One system that satisfies all the requirements on a paradoxical
magnetic coolant consists of a few ($4{\leq}N{\leq}14$)
super-paramagnetic particles of magnetite. Such particles are the key
constituents of ferrofluids, which have attained rapidly increasing
interest during the past few years \cite{Wang94,PRLxFFR}. Recently we
have shown that such systems exhibit intriguing phase transitions
between the ordered ring and chain phases and one disordered phase
\cite{JCPxTLS}. We also pointed out that self-assembly in these
systems could be used to store information \cite{SCIxMEM}.

In the following, we study the thermodynamic behavior of a
six-particle system, where the chain and the ring are the dominant
stable structural isomers. We chose this particular system as it
allows a simple discussion of the two thermodynamical features
characteristic of isomer transitions in finite systems, namely the
isomeric phase space and the transition probability that is linked to
the transition time. It is true that such ring and chain isomers are
also stable at much larger sizes. Nevertheless, the thermodynamical
behavior becomes more complex in larger systems, where the number of
relevant structural isomers grows rapidly with increasing $N$.

Here we calculate the entropy and the dependence of temperature on an
external magnetic field at constant entropy
$({\partial}T/{\partial}B)_S$ for a model system of six magnetite
particles with a radius $\sigma = 50$~{\AA} and a permanent magnetic
moment of $\mu = 2.63{\times}10^3~\mu_{\rm B}$. The total potential
energy $U$ of this system is given by \cite{PRLxFFR}
\begin{eqnarray}
U & = & \sum_{i<j}^{N} \left\{ (\mu_0^2/r_{ij}^3)
         \left[ \hat{\mu}_i\cdot\hat{\mu}_j
            - 3(\hat{\mu}_i\cdot\hat{r}_{ij})
               (\hat{\mu}_j\cdot\hat{r}_{ij})
         \right]  \right.  \\
\label{Eq1}
   &   & \left. + \epsilon
         \left[ e^{\left(-\frac{r_{ij}-\sigma}{\rho} \right)}
              - e^{\left(-\frac{r_{ij}-\sigma}{2 \rho}\right)}
  \right] \right\} + \sum_{i=1}^{N} \mu_z^i \cdot B_{\rm ext}\; .
  \nonumber
\end{eqnarray}
The pairwise interaction energy is given by the dipole-dipole and a
non-magnetic interaction energy. The latter is dominated by a
repulsion between the spherical particles, but contains also a weak
attractive part due the van der Waals interaction. It is modelled 
by the above Morse-type potential with 
parameters\cite{scaling} 
$\epsilon=15.1$~$\mu$eV and $\rho=2.5$~{\AA}.
The second sum reflects the interaction between the magnetite
particles and the external magnetic field $B_{\rm ext}$ that is
aligned with the $z$ axis. 

We would like to point out that the finite ferrofluid systems undergo
all the intriguing transitions described below occur independent of
these parameter values. Our particular choice has been taken to bring
the transition into an experimentally accessible and interesting
region \cite{parameters}.

\begin{figure}[htb]
\centerline{\psfig{figure=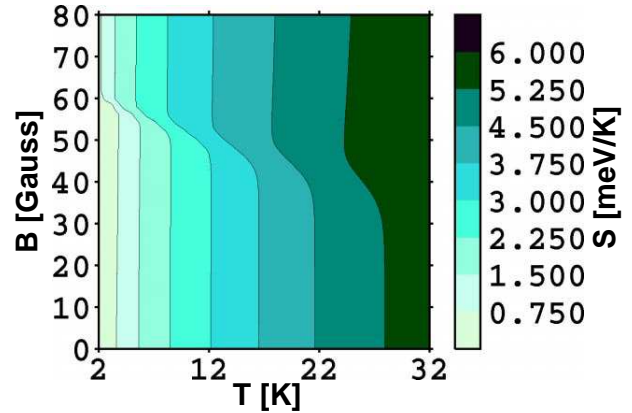,height=2.3in,angle=90}}
\caption{
Contour plot of the entropy $S$ as a function of the external magnetic
field $B_{\rm ext}$ and temperature $T$. The model system discussed
here shows a temperature decrease of up to few degrees Kelvin as the
field $B$ is increased at constant entropy.
}
\label{Fig1}
\end{figure}

\begin{figure}[htb]
\centerline{\psfig{figure=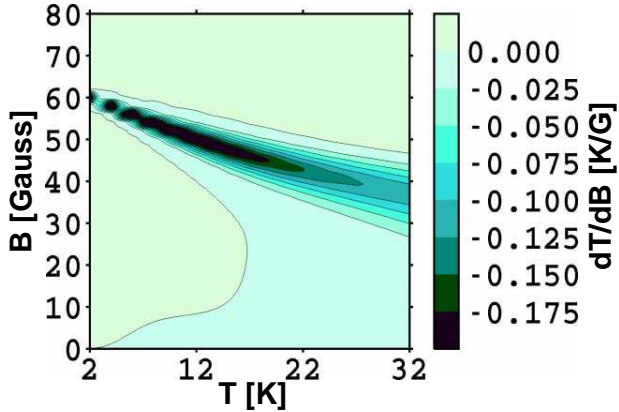,height=2.3in,angle=90}}
\caption{
Contour plot of $({\partial}T/{\partial}B)_{S}$ as a function the
external magnetic field $B_{\rm ext}$ and temperature $T$.
}
\label{Fig2}
\end{figure}

All thermodynamic quantities can be derived from the canonical
partition function $Z(B,T)$ by appropriate differentiation. We
determined $Z$ using the Metropolis Monte Carlo method
\cite{Metropolis}, which we combined with a special type of optimized
data analysis \cite{Ferrenberg} to calculate all thermodynamic
properties as functions of the temperature $T$ and the magnetic field
$B_{\rm ext}$ as external variables \cite{JCPxTLS}. The entropy $S$
is given by
\begin{equation}
   S=k_{\rm B} \left(\ln(Z)-\beta \frac{\partial
   \ln(Z)}{\partial \beta}\right) \; ,
\end{equation}
where $\beta=1/k_{\rm B}T$. The fundamental thermodynamic expression
${\rm d}E = T {\rm d}S - \mu_z {\rm d} B$ yields immediately the
Maxwell relation
\begin{equation}
  \left(\frac{\partial T}{\partial B}\right)_S = 
       \frac
       { -\left( \partial S/\partial B \right)_T } 
       {  \left( \partial S/\partial T \right)_B }
\end{equation}
that describes the temperature response to the external field in
isentropic processes. We calculate this quantity using the
expectation values of the potential energy $U$ and the $z$-component
of the magnetic moment $\mu_z$ of the whole system as
\begin{equation}
    \left( \frac{\partial T}{\partial B} \right)_S = - \beta \;
    \frac{
     \langle U \mu_z \rangle - \langle U \rangle \langle \mu_z \rangle
     }{\frac{6}{2} N k_B+k_B \beta^2 (\langle U^2 \rangle
     - \langle U \rangle^2) }\; .
\label{final}
\end{equation}
Our results for $S$ and $\left( \frac{\partial T}{\partial B}
\right)_S$ as a function of $B_{\rm ext}$ and $T$ are represented by
contour plots in Fig.~\ref{Fig1} and Fig.~\ref{Fig2}, 
respectively\cite{lowT}.

The steps in the isentropes displayed in Fig.~\ref{Fig1} indicate a
temperature decrease with increasing magnetic field at constant
entropy. For example, an increase of the field from 45 to 55 Gauss
cools the system down from $T=12$~K to $T{\approx}8$~K. The narrow region
in the $B_{\rm ext}-T$ space, where these kinks occur, separates the
chain and ring phases. It is only in this narrow region of the
$B_{\rm ext}-T$ space that $(\partial T/\partial B)_S$ shows a
nonzero value and hence a potential for magnetic cooling, indicated in
Fig.~\ref{Fig2}. A closer inspection of Fig.~\ref{Fig1} shows that
the $S={\rm const.}$ lines change their slope in the chain phase at
high fields, which also corresponds to a positive value of $(\partial
T/\partial B)_S$ in Fig.~\ref{Fig2}. This behavior is simply related
to the fact that chains behave like a {\sl conventional} magnetic
system, since their magnetic and structural order increases with
increasing external magnetic field. In the six-particle system
discussed above, the conventional cooling mechanism by isentropic
demagnetization in high fields is about one order of magnitude less
important than the paradoxical magnetic cooling that is related to
the ring-chain transition.

Let us now discuss some details related to a possible realization of
the system. In our simulations we disregarded the internal degrees of
freedom of the magnetic particles and the colloidal suspension.
Obviously, the cooling efficiency of this particular composite system
is highest when the magnetic degrees of freedom of the particles
dominate. A potential candidate paradoxical cooling system is a
dilute gas consisting of both magnetic particles, which could
aggregate, and non-magnetic particles of similar size. The
actual thermal transition is initiated by vibrational excitations
that lead either to the breakup of rings into chains or a reconnection
of chain ends to a ring.

Let us now consider a mixture of chains, rings, and other isomers
exposed to an oscillating magnetic field. Whereas the breakup energy
increases as the neighboring dipoles gradually lign up in the larger
rings, the vibrational excitation spectrum gets dense especially in
the range of low frequencies that drop with $1/N^2$. Hence it is
preferentially the larger rings that are broken up to chains by a
low-frequency external field. Also the chain structures experience an
analogous softening of the vibrational spectrum with the factor of
$1/N^2$. Yet due to the larger phase space volume, the likelihood of 
two chain ends to reconnect to a ring decreases drastically in larger 
systems.

The chain ends (which are to meet for a chain-to-ring transition) not
only move more slowly with larger $N$ but have to explore a much
larger space in which the other end can be. The time required for the
two ends to reconnect rises dramatically with $N$, so that
chain-to-ring transitions can not be realized in a reasonable time in
an ensemble of long chains. 

In a realistic experiment, we also have to consider that the average
size of the ferrofluid aggregates depends not only on the initial
conditions, but increases as a function of time due to further
aggregation. To prevent this gradual size increase, we propose to
expose the ferrofluid suspension to a low frequency magnetic field
that should preferentially break apart the larger aggregates and to
stabilize the mixture of prevalent isomers in the region of five to
seven particles per aggregate.

Paradoxical magnetic cooling is by no means restricted to the model
system discussed here. We believe that the same effect should also
occur in other nanostructures, such as transition metal clusters, and
even in bulk matter consisting of finite-size substructures. We hope
that our results will stimulate a search for experimentally realizable
systems that may find application in the fascinating field of
ultra-low temperature physics.

This work was partly supported by the Office of Naval Research
and DARPA under Grant No. N00014-99-1-0252. 




\begin{references}
\bibitem{parameters} Particles of different materials and sizes 
         can be used as constituents of ferrofluids. 
         The present choice of parameters, different from the choice
         in Ref.\ \protect\onlinecite{PRLxFFR}, represents a system 
         with a convenient transition temperature.
\bibitem{scaling} We note that in the present system of magnetic 
    particles, the temperature scales with the
    magnetic moment of the particles.
\bibitem{lowT} We chose $T=2$~K as the lower temperature limit for
    computational reasons that are related to the slow equilibration of a
    cold system.
\end{references}
\end{document}